\documentclass{IEEEtran4PSCC}

\usepackage{cite}

\ifCLASSINFOpdf
\usepackage[pdftex]{graphicx}
\else
\fi
\usepackage[cmex10]{amsmath}
\usepackage{array}
\usepackage{amssymb}
\usepackage{booktabs,colortbl}
\usepackage{color}
\usepackage{eurosym}
\usepackage{caption}
\usepackage{subcaption}

\newcommand{\Dtk}{\Delta t}

\newcommand{\ds}{\{\Pgsch{k}\}_{k \in \mathcal{K}}}

\newcommand{\CP}{\mathcal{P}_k}
\newcommand{\CE}{\mathcal{E}_k}
\newcommand{\probP}{\pi^{P}}
\newcommand{\probE}{\pi^{E}}

\newcommand{\ex}{\mathbb{E}}

\newcommand{\mun}{\mu_n}

\newcommand{\exPsp}[1]{\hat{p}_s^+(#1)}
\newcommand{\exPsm}[1]{\hat{p}_s^-(#1)}

\newcommand{\expPl}[1]{\hat{p}_{l}(#1)}
\newcommand{\expPs}[1]{\hat{p}_{s}(#1)}
\newcommand{\expEs}[1]{\hat{e}_{s}(#1)}

\newcommand{\Psmin}{\underline{p}_s}
\newcommand{\Psmax}{\overline{p}_s}
\newcommand{\Esmin}{\underline{e}_s}
\newcommand{\Esmax}{\overline{e}_s}

\newcommand{\Pgsch}[1]{\tilde{p}_g(#1)}

\newcommand{\Pgp}[1]{\tilde{p}_g^+(#1)}
\newcommand{\Pgm}[1]{\tilde{p}_g^-(#1)}

\newcommand{\kstart}{k_{s}}

\newcommand{\rv}[1]{\mathsf{#1}}

\newcommand{\nonl}{\renewcommand{\nl}{\let\nl\oldnl}}

\DeclareCaptionFont{tiny}{\tiny}

\begin{document}
%
\title{Storage Scheduling with Stochastic Uncertainties: Feasibility and Cost of Imbalances}

\author{
\IEEEauthorblockN{Riccardo Remo Appino, Jorge \'Angel Gonz\'alez Ordiano, Ralf Mikut, Veit Hagenmeyer, Timm Faulwasser}
\IEEEauthorblockA{Institute for Automation and Applied Informatics, Karlsruhe Institute of Technology \\
Karlsruhe, Germany\\
\{riccardo.appino, jorge.ordiano, ralf.mikut, veit.hagenmeyer, timm.faulwasser\}@kit.edu}
}


\maketitle

\begin{abstract}
Dispatchability of renewable energy sources and inflexible loads can be achieved using a volatility-compensating energy storage. 
However, as the future power outputs of the inflexible devices are uncertain, the computation of a dispatch schedule for such aggregated systems is non-trivial.
In the present paper, we propose a novel scheduling method that enforces the feasibility of the dispatch schedule with a pre-determined probability based on a description of the operation of the system as a two-stage decision process. 
Thereby, a crucial point is the use of probabilistic forecasts, in terms of cumulative density function, of the inflexible energy consumption/production profile. 
Then, for the sake of comparison, we introduce a second scheduling method based on state-of-the-art scenario optimization, where, unlike the proposed method, the focus is on the minimization of the expected final cost.
We draw upon simulations based on real consumption and production data to compare the methods and illustrate our findings.
\end{abstract}


\thanksto{The authors gratefully acknowledge funding by the German Federal Ministry of Education and Research (BMBF) within the Kopernikus Project ENSURE ‘New ENergy grid StructURes for the German Energiewende’. Moreover, JAGO, RM, VH, and TF  acknowledge support by the Helmholtz Association under the Joint Initiative ``Energy System 2050– A Contribution of the Research Field Energy''.
TF acknowledges further support from the Baden-W\"urttemberg Stiftung under the Elite Program for Postdocs.}

\vspace{-0.3cm}
\section{Introduction}
\label{sec:intro}

The increasing amount of relatively inflexible, volatile, and unpredictable distributed generation from renewable energy resources calls for more flexibility at each level of the electrical energy chain. This flexibility is mandatory for maintaining power balance and for an efficient operation of power systems. 
Energy Storage Systems (ESSs) are seen as a particularly promising source of flexibility. 
ESSs could be used, for example, to compensate for volatile generation and consumption, thus enabling the dispatch of power output from inflexible generation and/or demand according to a pre-computed schedule \cite{Sossan16a}. 
Following the notion introduced in \cite{Sossan16a}, we employ the term \textit{dispatchable feeder} to refer to a grid-connected power system composed of an ESS and inflexible generation/demand whose power output is regulated according to a pre-computed schedule. 
We denote this schedule as \textit{Dispatch Schedule} (DiS).

Several scheduling and control schemes have been proposed in the literature to operate dispatchable feeders, e.g. \cite{Sossan16a,Stai17,Lampropoulos15}.
The computation of the DiS constitutes one of the major challenges in all these works. 
In fact, the exact upcoming inflexible generation/demand is unknown at the moment in which the DiS has to be computed. 
Even if forecasts for this inflexible generation/demand are available, they are still prone to errors. 
Therefore, the future power outputs are known, at best, in terms of random variables \cite{Zhang14}.
These variables are often not normally distributed \cite{Salameh95} and it is generally difficult to describe their inherent dependency in an explicit mathematical form. 
For these reasons, most of the well-known techniques used to deal with random disturbances, e.g. \cite{Farina16}, are inapplicable. 
Therefore, two-stage stochastic programming with scenario selection is often applied in computation of DiSs, cf. \cite{Olivares15,Ding12,Garcia08,Vrakopoulou13,Stai17}.
Alternatively, multi-stage robust optimization can be used \cite{Fabietti16}, but might be conservative. 
To overcome this issue, our preceding paper \cite{Appino18a} proposes stochastic robust optimization ensuring trackability of the DiS with a given probability. The underlying idea is the use of probabilistic forecasts of the energy profile, thus implicitly including the temporal correlations of the future power outputs. 
The present paper extends this approach.
      
The contributions are as follows: 
First, we formulate the scheduling problem as a two-stage decision process. 
Then, taking this model as a starting point, we extend the stochastic robust formulation presented in \cite{Appino18a}, including a better exploitation of the probabilistic forecasts; 
i.e., while \cite{Appino18a} considered quantile-based energy forecasts, herein we work directly with Cumulative Density Functions (CDF).
Moreover, we investigate constraint softening to the end of avoiding infeasibility. 
Finally, we compare the proposed approach to a scenario-based scheduling method similar to \cite{Stai17,Fabietti16}, relying on scenarios generated via probabilistic forecasts. 
Our results show that the proposed scheduling problem based on probabilistic forecasts of the energy profile outperforms both scheduling based on deterministic forecast and based on scenario forecasts.

The remainder of the present paper is organized as follows: Section \ref{sec:problem_state} covers the problem setup; Section \ref{sec:security_level} presents the main contribution, i.e. the proposed methodology to tackle the stochastic scheduling problem using the CDF of the forecast quantities; Section \ref{sec:minimal_cost} describes scheduling using scenario-based optimization; Section \ref{sec:results} reports simulation results.

\section{Problem Statement}
\label{sec:problem_state}

\subsection{System Description}

\begin{figure}[t]
\vspace{-0.2cm}
\centering
\includegraphics[width=0.46\textwidth]{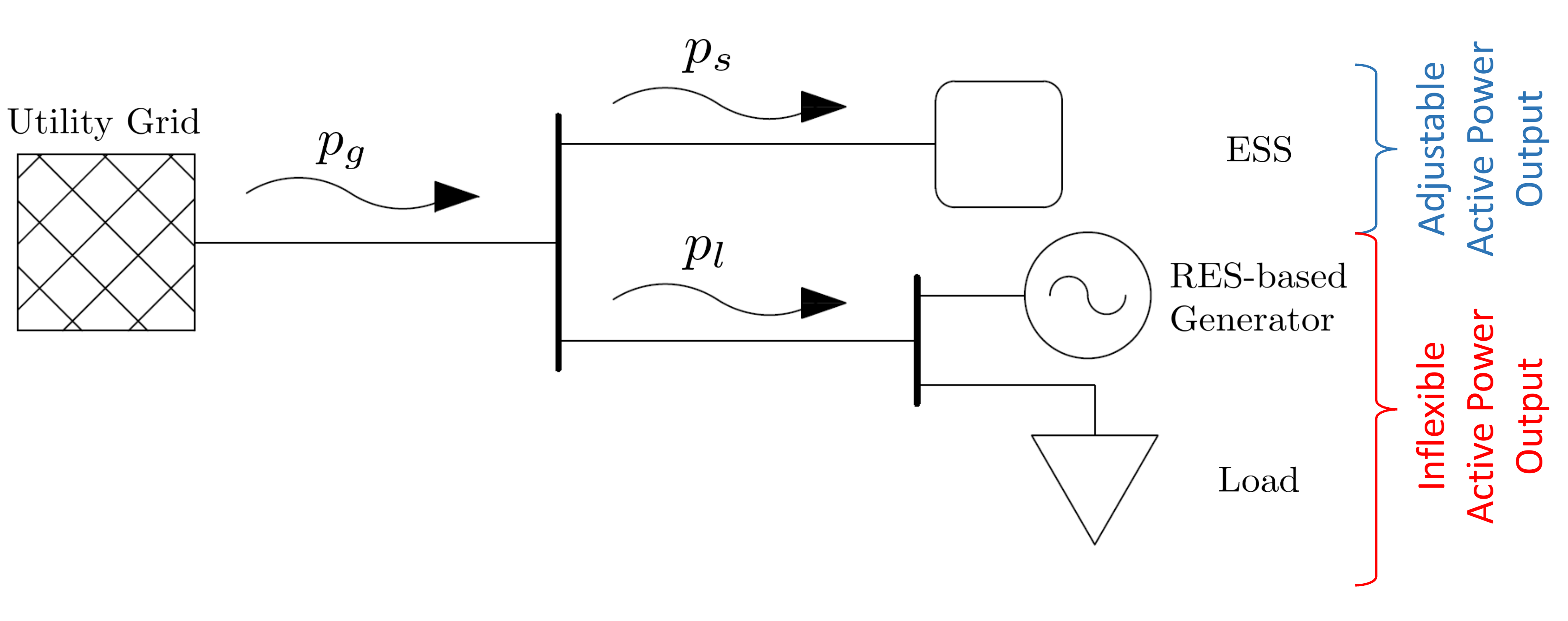}
\caption{Schematic diagram of a generic \textit{dispatchable feeder} (modified from \cite{Appino18a}). \label{fig:Scheme} 
}
\vspace{-0.4cm}
\end{figure}

Similar to \cite{Appino18a}, the present paper addresses the optimal operation of a dispatchable feeder, cf. Figure \ref{fig:Scheme}. 
The dispatchable feeder is operated such that the active power exchange with the utility grid, $p_g$, follows a pre-computed \textit{Dispatch Schedule} (DiS). 
The time window covered by the DiS is called the \textit{scheduling horizon}. 
In the following, we adopt a discrete time notation and divide the scheduling horizon into $N_d \in \mathbb{N}$ \textit{dispatch intervals} of equal duration $\Dtk$. 
We enumerate the dispatch intervals with the index $k \in \mathcal{K}= [\kstart, \kstart+N_d] \subset \mathbb{N}$ and indicate with $\Pgsch{k}$ the DiS at interval $k$. 
We refer to the sequence of $\Pgsch{k}$ over the scheduling horizon, i.e. the DiS, as $\ds$.
The DiS is computed at $k_0 < \kstart$, before the scheduling horizon.

The operation of the dispatchable feeder can be described using a two-stage decision process. At the first stage $\ds$ is computed. 
In the second stage, the ESS power output at $k$, $p_s(k)$, is determined.
In the following, we describe the details of these two stages, starting from the second one.

The second stage of the decision process takes place \textit{after} the aggregated power output of the inflexible elements is known. 
The notation $p_l(k)$ indicates the value of this power output at $k$, with negative values representing generation.
The ESS is used to compensate the volatility of the inflexible elements, as in \cite{Citro11}. 
Its power output, $p_s(k)$, is therefore chosen to meet the scheduled power exchange with the grid
\begin{equation}
p_s(k) = \Pgsch{k} - p_l(k). \label{eq:real_pow_bal}
\end{equation}
However, $p_s(k)$ cannot assume arbitrary values, being limited by the capacity and capability constraints of the ESS
\begin{subequations}
\label{eq:det_ess_limit}
\begin{align}
\Psmin \leq {p}_s(k) \leq \Psmax, \label{eq:det_pow_limit}\\
\Esmin \leq {e}_s(k) \leq \Esmax, \label{eq:det_en_limit}
\end{align}
\end{subequations}
with
\begin{subequations}
\label{eq:complem_ess_limit}
\begin{multline}
e_s(k+1) = e_s(k) + \left( (1-\mun)p_s^+(k) + (1+\mun)p_s^-(k) \right) \Dtk, \\e_s(k_0) = e_s^{k_0},\label{eq:storage_dinamic_det}
\end{multline}
\begin{equation}
[p_s^+(k),p_s^-(k)] \in \mathcal{F}_{d}(p_s(k)).
\end{equation}
\end{subequations}
Here, $\Psmin$ and $\Psmax$ denote the minimum and maximum power output, $\Esmin$ and $\Esmax$ denote the minimum and maximum capacity.
Finally, $e_s(k)$ is the stored energy at $k$, and $e_s^{k_0}$ is the initial state of charge of the storage. 
The conversion losses are modeled by $\mun \in [0,1]$, together with a discrimination between the different directions of $p_s(k)$, i.e. $p_s^+(k)$ and $p_s^-(k)$.
To describe this discrimination, we introduce the set 
\begin{align*}
\mathcal{F}_{d}(p) := \Big\{ [p^+,p^-]^\top \in \mathbb{R}^{2} \,|&\,\, p^- \cdot p^+ = 0, \, p^+ \geq p, \, p^- \leq p, \\& p^+ \geq 0, \,  p^- \leq 0. \Big\}.
\end{align*} 
Notice that, given a certain $\Pgsch{k}$, the value of $p_s(k)$ computed using \eqref{eq:real_pow_bal} might violate constraints \eqref{eq:det_ess_limit} for some values of $p_l(k)$. 
We assume that, when this is the case, the power balance is maintained by deviating the power exchange with the utility grid from the DiS. 
These deviations are often referred to as \textit{imbalances} \cite{Morales13}. 
We denote imbalances with $\Delta p_g(k)$. The total power exchange with the grid at instant $k$ is therefore
\begin{equation}
p_g(k) = \Pgsch{k} + \Delta p_g(k).
\end{equation} 
Thus, the active power balance at the second stage is
\begin{equation}
p_s(k) = \Pgsch{k} + \Delta p_g(k) - p_l(k). \label{eq:real_pow_bal_dev}
\end{equation}
According to this model, $\Delta p_g(k)$ is also determined at the second-stage, once the value of $p_l(k)$ is known. Specifically, $p_s(k)$ and $\Delta p_g(k)$ follow from the optimization
\begin{align}
\label{eq:delta_pg_opt_problem}
\min_{p_s(k),\Delta p_g(k)} \Delta p_g(k) \quad
\text{s.t. \,\,}  \eqref{eq:det_ess_limit}, \eqref{eq:complem_ess_limit}, \eqref{eq:real_pow_bal_dev}.
\end{align}

Recall that $\Pgsch{k}$ is computed at $k_0$, \textit{before} the value of $p_l(k)$ is known. This computation is the first stage of the decision process.
At time $k_0$, a probabilistic forecast is the only available 
information on the inflexible power output.
This forecast is described by a random variable, $\rv{P}_l(k)$, whose realization is $p_l(k)$. 
Thus, the power balance at the \textit{first stage} is
\begin{equation}
\rv{P}_s(k) = \Pgsch{k} + \rv{\Delta P}_g(k) - \rv{P}_l(k). \label{eq:gen_power_bal}
\end{equation}
Note that $\rv{\Delta P}_g$ and $\rv{P}_s(k)$ are random variables, as their realizations depend on $p_l(k)$, unknown at this stage. 
As the ESS power output $\rv{P}_s(k)$ is a stochastic quantity, so is the energy stored at time instant $k$, $\rv{E}_s(k)$.
In the following, we employ a model with approximated ESS conversion losses to describe the dynamics of $\rv{E}_s(k)$, similar to \cite{Appino18a}. 
Specifically, we decouple deterministic and stochastic variables via 
\begin{subequations}
\label{eq:exp_value_notation}
\begin{align}
\expPl{k} &= \ex[\rv{P}_l(k)], \\
\rv{P}_l(k) &= \expPl{k} + \rv{\Delta P}_l(k),\\
\rv{P}_s(k) &= \expPs{k} + \rv{\Delta P}_s(k), \\
\rv{E}_s(k) &= \expEs{k_0} + \rv{\Delta E}_s(k).
\end{align}
\end{subequations}
Requiring 
\begin{equation}
\label{eq:exp_ess_power}
\expPs{k} = \Pgsch{k} - \expPl{k},
\end{equation}
the equality
\begin{equation}
\rv{\Delta P}_s(k) = \rv{\Delta P}_g(k) - \rv{\Delta P}_l(k), \label{eq:deviation_p_bal}
\end{equation}
follows from \eqref{eq:gen_power_bal}.
Relying on this notation, we describe the dynamics of the expected value of $\rv{E}_s(k)$ using \eqref{eq:storage_dinamic_det}, i.e. 
\begin{subequations}
\label{eq:exp_ess_energy}
\begin{multline}
\expEs{k+1} = \expEs{k} + \left( (1-\mun)\exPsp{k} + (1+\mun)\exPsm{k} \right) \Dtk, \\
\expEs{k_0} = e_s^{k_0},
\end{multline}
\begin{equation}
[\exPsp{k},\exPsm{k}] \in \mathcal{F}_{d}(\expPs{k}).
\end{equation}
\end{subequations}
For the stochastic variable $\rv{\Delta E}_s(k)$ we employ an approximated loss-less model\footnote{This choice leads to a tractable stochastic model while introducing a negligible error, see \cite{Appino18a} for details.} ($\mu_n=0$)
\begin{align*}
\rv{\Delta E}_s(k+1) &= \rv{\Delta E}_s(k) + \rv{\Delta P}_s(k)\Dtk \nonumber \\
&= \rv{\Delta E}_s(k) + \left(\rv{\Delta P}_g(k) - \rv{\Delta P}_l(k)\right)\Dtk \nonumber \\
&= \sum_{i=k_0}^{k} \left(\rv{\Delta P}_g(i) - \rv{\Delta P}_l(i)\right)\Dtk,
\end{align*}
with initial condition $\rv{\Delta E}_s(k_0) = 0$ (the state of charge at $k=k_0$ is known).
Using 
\begin{align*}
\rv{\Delta E}_g(k) = \sum_{i=k_0}^{k-1} \rv{\Delta P}_g(i) \Dtk,\\
 \rv{\Delta E}_l(k) = \sum_{i=k_0}^{k-1} \rv{\Delta P}_l(i) \Dtk,
\end{align*}
we obtain
\begin{equation}
\rv{\Delta E}_s(k) = \rv{\Delta E}_g(k) - \rv{\Delta E}_l(k) \label{eq:deviation_e_bal}.
\end{equation}
Note that $\rv{\Delta P}_l(k)$ and $\rv{\Delta P}_l(j)$ with $k\not= j$ are correlated, i.e. they are \textit{not} independent random variables. 
Therefore, it is in general difficult to compute $\rv{\Delta E}_l(k)$. 
In the following, we address this issue in two different ways: we utilize probabilistic forecasts of this variable in Section \ref{sec:security_level}, and we consider scenario forecasting (i.e. ensemble forecasts) in Section \ref{sec:minimal_cost}.

Furthermore, the capacity and capability limits of the storage \eqref{eq:det_ess_limit} at the first decision stage are, with a slight abuse of notation,  
\begin{subequations} \label{eq:stoch_inex_ess_limit}
\begin{align}
\Psmin \leq \rv{P}_s(k) \leq \Psmax, \label{eq:inex_pow_limit}\\
\Esmin \leq \rv{E}_s(k) \leq \Esmax, \label{eq:inex_en_limit}
\end{align}
meaning that \eqref{eq:stoch_inex_ess_limit} should hold for each realization of $\rv{P}_s(k)$ and $\rv{E}_s(k)$.
\end{subequations}
Given \eqref{eq:exp_value_notation}, \eqref{eq:deviation_p_bal}, and \eqref{eq:deviation_e_bal}, the first stage constraint \eqref{eq:stoch_inex_ess_limit} 
can be described with
\begin{subequations}
\label{eq:inex_ess_limit_pl}
{\begin{gather}
\Psmin \leq \Pgsch{k} - \expPl{k} + \rv{\Delta P}_g(k) - \rv{\Delta P}_l(k) \leq \Psmax,\label{eq:inex_pow_limit_pl}\\
\Esmin \leq \expEs{k} + \rv{\Delta E}_g(k) - \rv{\Delta E}_l(k) \leq \Esmax.\label{eq:inex_en_limit_pl}
\end{gather}}
\end{subequations}
\subsection{Scheduling Requirements}
The requirements of DiS computation and optimization can be split into two major categories:
\begin{itemize}
\item \textit{Operational requirements}, i.e. the explicit requirements of the DiS such as peak shaving, load leveling and price-dependent load shifting.  
\item \textit{Tracking requirements}, i.e. the implicit requirement of tracking the DiS by means of an underlying intra-schedule controller.
\end{itemize}
The presence of these two, possibly conflicting, categories of requirements makes the computation of the DiS particularly challenging. 
In the following, we discuss how these requirements reflect the proposed two-stage decision process.

We translate the operational requirements into an appropriate cost function, where a lower cost indicates improved satisfaction of requirements. 
As the DiS is deterministic at the first stage, its associated cost is also \textit{deterministic}.
Here, we consider a quadratic cost function of the power exchange scheduled via the DiS 
\begin{align}
\label{eq:cost_f}
\begin{split}
C(\Pgp{k},\Pgm{k}) = &c_{1}^+(k)(\Pgp{k})^2+c_{2}^+(k) \Pgp{k}\\+&c_{1}^-(k)(\Pgm{k})^2+c_{2}^-(k) \Pgm{k}, 
\end{split}
\end{align} 
where $c_{1}^+(k)$, $c_{1}^-(k)$, $c_{2}^+(k)$, $c_{2}^-(k)$ are time-varying cost coefficients and $\Pgp{k}$ and $\Pgm{k}$ represent the different directions of $\Pgsch{k}$, subject to
\begin{equation}
\label{eq:Pg_dir}
	[\Pgp{k},\Pgm{k}] \in \mathcal{F}_{d}(\Pgsch{k}).
\end{equation}
By an appropriate choice of cost coefficients, cost function \eqref{eq:cost_f} can be used to achieve load leveling and/or price-based load shifting.
 
The tracking requirements, instead, involve minimizing and/or constraining the imbalances, $\Delta p_g(k)$. 
We remark that, at the first stage, $\Delta p_g(k)$ is unknown and represented by the random variable $\rv{\Delta P}_g(k)$.
Thus, the tracking requirements involve dealing with uncertainty.
In this paper, we consider two separate tracking requirements inherently:
(i) limiting the number of imbalances, and
(ii) minimizing the cost of imbalances. 
We approach these requirements individually, with different techniques.
We tackle the first one by applying joint chance constraints to the scheduling problem (Section \ref{sec:security_level}).
We consider the second one by using scenario-based optimization (Section \ref{sec:minimal_cost}).

\section{Enforcing a Given Security Level}
\label{sec:security_level}

Now, we formulate a scheduling problem to minimize the cost of the DiS while considering the requirement of limiting the number of imbalances. 
Specifically, we consider a DiS that can be tracked with a given probability at each time instant. 
We refer to this probability, $(1 - \varepsilon)$, as \textit{security level}. 
Given the model of the system described in the previous section, the absence of deviations corresponds to $\Delta p_g(k)=0$ and, therefore, to $\text{P}[\rv{\Delta P}_g(k)=0]=1$ and $\text{P}[\rv{\Delta E}_g(k)=0]=1$.
Consequently, ensuring that the DiS at $k$, $\Pgsch{k}$, is met with at least security level $(1 - \varepsilon)$ equals to  
\begin{align}
&\text{P}[ \CP \cap \CE] \geq (1 - \varepsilon),\label{eq:joint_ch_constr_both}
\end{align}
where the events $\CP$ and $\CE$ are related to satisfying the ESS constraints \eqref{eq:inex_ess_limit_pl} without any deviation from the schedule, i.e.  
\begin{align*}
\CP &= \left \lbrace \Psmin \leq \Pgsch{k} - \expPl{k} - \rv{\Delta P}_l(k) \leq \Psmax \right \rbrace, \nonumber \\
\CE &= \left \lbrace \Esmin \leq \expEs{k} - \rv{\Delta E}_l(k) \leq \Esmax \right\rbrace. 
\end{align*}
Instead of tackling constraint \eqref{eq:joint_ch_constr_both} directly, we first formulate and analyze separate joint chance constraints for power and energy \cite{Appino18a}:
\begin{subequations}
\label{eq:joint_ch_constr_P_and_E}
\begin{align}
&\text{P}[\CP ] \geq (1 - \varepsilon_P)\text{,} \label{eq:joint_ch_constr_P}\\ 
&\text{P}[\CE] \geq (1 - \varepsilon_E)\text{.} \label{eq:joint_ch_constr_E}
\end{align}
\end{subequations}
Then, we require the DiS to be robust against worst-case realization of the uncontrolled power output $\rv{P}_l(k)$,\footnote{Throughout all simulations reported in Section \ref{sec:results}, this worst-case choice for $\varepsilon_P$ has not led to infeasibility of the scheduling problem.} i.e. $(1 -\varepsilon_P) \simeq 1$, which implies
$\text{P}[\CP] \simeq 1$ and 
$\text{P}[\CP \cup \CE] \simeq 1$. 
Thus
$
\text{P}[ \CP \cap \CE] = \text{P}[\CP] + \text{P}[\CE] - \text{P}[\CP \cup \CE] \simeq \text{P}[\CE],
$
and
$
\text{P}[ \CP \cap \CE] \geq (1 - \varepsilon_E),
$
meaning that \eqref{eq:joint_ch_constr_both} is respected if  \eqref{eq:joint_ch_constr_P_and_E} is satisfied with $(1 -\varepsilon_P) \simeq 1$ and $(1 -\varepsilon_E) \geq (1 -\varepsilon)$.

Numerical tractability of the chance constraint \eqref{eq:joint_ch_constr_P} is achieved as in \cite{Appino18a}.
Specifically, rewriting the event $\CP$ as
\begin{align*}
\CP &= \left\lbrace \Pgsch{k} - \Psmax \leq \rv{P}_l(k) \leq \Pgsch{k} - \Psmin \right\rbrace,
\end{align*} 
and given the probabilistic forecast for $\rv{P}_l(k)$
\begin{equation}
\text{P} \Big[ \rv{P}_l(k) \in [\underline{p}_{l,\probP}(k),\overline{p}_{l,\probP}(k)] \Big] = \probP = (1 - \varepsilon_P) \simeq 1, \label{eq:interval_Pl}
\end{equation}
constraint \eqref{eq:joint_ch_constr_P} is satisfied with $(1 - \varepsilon_P) \simeq 1$ if
\begin{equation}
\Pgsch{k} - \Psmax \leq \underline{p}_{l,\probP}(k) , \quad  \overline{p}_{l,\probP}(k) \leq \Pgsch{k} - \Psmin, \label{eq:p_costr_bot}
\end{equation}
i.e. if $\CP$ is verified for (approximately) the entire support of $\rv{P}_l(k)$.

With respect to the energy constraint \eqref{eq:joint_ch_constr_E}, instead, we use a different and improved technique in comparison to \cite{Appino18a}. 
Therein, the event $\CE$ is restated as
\begin{align*}
\CE = \{  \expEs{k} - \Esmax  \leq \rv{\Delta E}_l(k) \leq \expEs{k} - \Esmin \},
\end{align*} 
and \eqref{eq:joint_ch_constr_E} is achieved using a percentile approach 
\begin{align}
\label{eq:old_constraint_ref}
\expEs{k} - \Esmax  \leq \Delta \underline{e}_{l,\probE}(k) , \quad  \Delta \overline{e}_{l,\probE}(k) \leq \expEs{k} - \Esmin,
\end{align}
with interval forecasts built around the median
\begin{align*}  
	&\text{P} \Big[ \rv{\Delta E}_l(k) \in [\Delta \underline{e}_{l,\probE}(k),\Delta \overline{e}_{l,\probE}(k)] \Big] = \probE = (1 - \varepsilon_E).
\end{align*}
However, this method presents several limitations.
First, considering different---even if larger---intervals with the same probability of containing the realizations of $\rv{\Delta E}_l(k)$ might lead to a reduction of the cost of the DiS.
Second, the scheduling problem might be infeasible for an arbitrarily large value of $(1 - \varepsilon_E)$. 
This second issue is particularly relevant in the case of long-term scheduling, as the support of $\rv{\Delta E}_l(k)$, growing with $k$, might become very large in comparison to the available storage capacity.  
 
Recalling that \eqref{eq:joint_ch_constr_E} is equivalent to
\begin{align}
&F_{\rv{\Delta E}_l(k)}\left( \expEs{k} - \Esmin \right) - F_{\rv{\Delta E}_l(k)}\left(\expEs{k} - \Esmax \right)\geq (1 - \varepsilon_E), \label{eq:e_costr_bot}
\end{align}
we propose in the present paper to use probabilistic forecasts for the CDF of $\rv{\Delta E}_l(k)$, $F_{\rv{\Delta E}_l(k)}(\Delta e_l)$,
to alleviate the issues of \eqref{eq:old_constraint_ref}.\footnote{We recall that, given a random variable $\rv{Z} \in \mathrm{L}^2(\Omega, \mu; \mathbb{R})$, the CDF is a function $F_{\rv{Z}}(z): \Omega \rightarrow [0,1]$ such that $F_{\rv{Z}}(z):=\text{P}[\rv{Z} \leq z]$. If $F_{\rv{Z}}(z)$ is known, the inequalities $F_{\rv{Z}}(a) - F_{\rv{Z}}(b) \geq (1 - \varepsilon)$, $\overline{z} \leq a$, and $\underline{z} \geq b$ are a deterministic reformulation of the joint chance constraint $\text{P}[\underline{z} \leq \rv{Z} \leq \overline{z}] \geq (1 - \varepsilon)$,
\cite{Miller65}. 
As $F_{\rv{Z}}(z)$ is increasing, all these inequalities are contemporarily verified if $F_{\rv{Z}}(\overline{z}) - F_{\rv{Z}}(\underline{z}) \geq (1 - \varepsilon)$ holds.
}

The proposed constraints formulation is similar, yet not equivalent, to the concept of stochastic robust optimization presented in \cite{King12}. 
In fact, similar to \cite{King12}, the satisfaction of joint chance constraint \eqref{eq:joint_ch_constr_P_and_E} is achieved by enforcing problem feasibility for convex compact subsets of the support of each random variable.
However, contrary to \cite{King12}, these subsets are chosen using full information on the distribution of the random variables   
and do not have to be symmetric w.r.t. the expected value. 
In this sense, the proposed constraints formulation extends the one based on interval forecasts presented in \cite{Appino18a}, avoiding the usage of fixed intervals. 

Moreover, the reformulation of the energy constraint \eqref{eq:e_costr_bot} 
enables to overcome the infeasibility problem with constraint softening \cite{Kerrigan00}. 
Specifically, \eqref{eq:e_costr_bot} can be replaced by
\begin{equation}
\label{eq:e_costr_bot_soft}
F_{\rv{\Delta E}_l(k)}\left(\hat{e}_s(k) - \Esmax \right) - F_{\rv{\Delta E}_l(k)}\left( \hat{e}_s(k) - \Esmin \right) + (1 - \varepsilon_E) \leq \epsilon(k),
\end{equation}
adding the penalty term $\alpha \cdot \epsilon(k)$ to the cost function of the scheduling problem, introduced later, with a sufficiently large value of $\alpha$ \cite{Kerrigan00}. 
This technique maximizes the probability of satisfying the energy constraint when it is not possible to guarantee the desired security level, making this approach particularly interesting for robust optimization. 

To sum up, we propose to compute the DiS solving 
\begin{align}
\label{eq:pfs_opt_problem}
\min_{
\begin{subarray}{c}
  \{\mathbf{p}_a(k)\}_{k\in \mathcal{K}'}, \\
  \{\epsilon(k)\}_{k\in \mathcal{K}'}
  \end{subarray}} \sum^{\kstart + N_d + N_f}_{k = \kstart} &\left( C\left(\Pgsch{k}\right) + \alpha \epsilon(k) \right)\\
\text{s.t. \,\,} & \eqref{eq:exp_ess_power},\eqref{eq:exp_ess_energy}, \eqref{eq:Pg_dir}, \eqref{eq:p_costr_bot}, \eqref{eq:e_costr_bot_soft} \nonumber
\end{align}
where the cost function $C$ is from \eqref{eq:cost_f} and $\mathcal{K}'= [\kstart, \kstart+N_d+N_f] \subset \mathbb{N}$. For each time step $k$, the stacked vector of decision variables is 
\begin{align*}
 \bold{p}_a(k)   := [\Pgp{k}, \Pgm{k}, \exPsp{k}, \exPsm{k}]^\top \in \mathbb{R}^4.
\end{align*}
Observe that Problem \eqref{eq:pfs_opt_problem} is non-convex, that there is no cost associated to the value of $\expEs{\kstart+N_{d}}$, and that a non-zero value for $\Delta e_g(k_s)$, i.e. a non-zero realization of $\rv{\Delta E}_g(k_s)$, might be included in \eqref{eq:e_costr_bot_soft}.
For a detailed description of how one can address these issues and how to choose the extended horizon $N_d + N_f$, we refer the reader to \cite[Remarks 1-3]{Appino18a}.
Finally, notice that no restrictive assumption on the distribution of either $\rv{P}_l(k)$ or $\rv{\Delta E}_l(k)$ is required in the reformulation of the joint chance constraints \eqref{eq:joint_ch_constr_P_and_E} into \eqref{eq:p_costr_bot} and \eqref{eq:e_costr_bot_soft}.

\section{Minimizing the Cost of Deviations}
\label{sec:minimal_cost}

In this section, we propose a scheduling formulation alternative to Problem \eqref{eq:pfs_opt_problem}, aiming to minimize the total expected operating cost, including the cost of the DiS and the expected cost of imbalances. 
This cost can be expressed as
\begin{equation}
C_{\text{tot}}(\Pgsch{k},\rv{\Delta P}_g(k)) = C(\Pgsch{k}) + \ex[C_{\text{dev}}(\rv{\Delta P}_g(k))].
\end{equation} 
As for the DiS cost, the cost of imbalances $C_{\text{dev}}(\Delta p_g(k))$ might reflect a real monetary cost in a two-stage market, c.f. \cite{Morales13}, or model different requirements on the imbalances. Differently from what presented in Section \ref{sec:security_level}, in this case the uncertainty does not only affect the constraints, but it enters the cost function too. 
Thus, it is not important to determine \textit{if} an imbalance will occur, but what would be \textit{its consequences} on the final cost. 
Likewise, it is important to know \textit{when} and \textit{with which magnitude} the imbalance will occur. 
To this end, given the difficulties of computing an explicit description of the correlation between $\rv{\Delta P}_l(k)$ at different intervals and the non-linearity of the ESS model, we adopt a scenario approach similar to \cite{Stai17,Fabietti16}. Each scenario $i$ describes a possible profile of the inflexible power output, i.e. $\{p_l^i(k)\}_{k\in \mathcal{K}}$, and has a certain weight $\omega_i$. Differently from \cite{Stai17,Fabietti16}, however, we generate the scenarios starting from probabilistic forecasts of $\rv{P}_l$, as described in Section \ref{sec:results}. 

For the sake of concise notation, we introduce 
\begin{align*}
 \bold{p}_s^i(k) &:= [p_s^{+,i}(k), p_s^{-,i}(k), \Delta p_g^{+,i}(k),\Delta p_g^{-,i}(k)]^\top \in \mathbb{R}^{4}, \\
 \bold{p}_b(k)   &:= [p_g^+(k), p_g^-(k), \bold{p}_s^1(k)^\top, \hdots, \bold{p}_s^{N_s}(k)^\top]^\top \in \mathbb{R}^{2+4N_s}, 
\end{align*}
where $N_s$ is the number of scenarios,
and the set
\begin{align*}
\mathcal{F}_{s}\Big(e_s(k)\Big) := \{ p_s(k) \in \mathbb{R} \,|&  \text{s.t. } \eqref{eq:det_pow_limit}, \eqref{eq:complem_ess_limit}\text{, and}\\ & \Esmin \leq {e}_s(k+1) \leq \Esmax \text{ hold}.\}.
\end{align*}  
The optimization problem based on scenario forecasts is 
\begin{align}
\label{eq:opt_problem}
\min_{\{\mathbf{p}_b(k)\}_{k\in \mathcal{K}'}} &\sum^{\kstart + N_d + N_f}_{k = \kstart} \left( C\left(\Pgsch{k}\right) + \sum^{N_s}_{i = 1} \omega_i C_{\text{dev}} \left( \Delta p_g^i(k) \right) \right)\\
\text{s.t. \,\, } & p_s^i(k) = \Pgsch{k} + \Delta p_g^i(k) - p_l^i(k) \,\, \forall k \in \mathcal{K}', \nonumber \\
& {p}_s^i(k) \in \mathcal{F}_{s}\Big(e_s(k)\Big) \,\, \forall k \in \mathcal{K}', \nonumber \\
& [\Pgp{k},\Pgm{k}] \in \mathcal{F}_{d}(\Pgsch{k}) \,\, \forall k \in \mathcal{K}', \nonumber \\
& [\Delta p_g^{+,i}(k),\Delta p_g^{-,i}(k)] \in \mathcal{F}_{d}(\Delta p_g^i(k)) \,\, \forall k \in \mathcal{K}'. \nonumber  
\end{align}

As described in Section \ref{sec:problem_state}, $\ds$ are first-stage decision variables, independent from the realization of the random variables, i.e. from $\{p_l^i(k)\}_{k\in \mathcal{K}}$.
On the other hand, the imbalances profile $\{\Delta p_g^i(k)\}_{k\in \mathcal{K}}$ is a sequence of second-stage scenario-based decision variables that depend both on the first-stage decision, $\ds$, and on the inflexible power output of the specific scenario, $\{p_l^i(k)\}_{k\in \mathcal{K}}$.
In Problem \eqref{eq:opt_problem} $\{\Delta p_g^i(k)\}_{k\in \mathcal{K}}$ is optimized for each scenario, with the implicit assumption that, once the DiS is applied, the imbalances can be planned with perfect knowledge of the profile $\{p_l(k)\}_{k\in \mathcal{K}}$.
As remarked in \cite{Fabietti16}, this is an approximation, since the imbalances are actually computed with the sole knowledge of the present values of $p_l(k)$ and $e_s(k)$, c.f. Problem \eqref{eq:delta_pg_opt_problem}.
Extending the horizon of Problem \eqref{eq:delta_pg_opt_problem} using short-term deterministic forecast, i.e. adding an additional model predictive control level to optimize the imbalances as in \cite{Appino18a}, might reduce the severity of this approximation. 
We include this additional controller in the following simulations. 
 
\section{Simulations and Results}
\label{sec:results}

\begin{table*}[!t]
\small
\renewcommand{\arraystretch}{1.1}
	\caption{Simulation results. Bold numbers highlight lowest cost.\label{tab:simulation_results}}
	\vspace{-0.2cm}
	\begin{center}
		\begin{tabular}{ l || c | c c | c c c c c c | c c}
		\hline &&&&&&&&&&& \vspace{-0.3cm}\\
		& DFS &\multicolumn{2}{c}{PFS (from \cite{Appino18a})} &&& \multicolumn{2}{c}{PFS} &&& \multicolumn{2}{c}{SFS} \\
		\hline &&&&&&&&&&& \vspace{-0.3cm} \\
		\hline &&&&&&&&&&& \vspace{-0.3cm} \\
			$(1 - \varepsilon_E)$ or $C1$/$C2$ & - &$\phantom{0}0.42$ &$\phantom{0}0.48$ &$\phantom{0}0.42$ &$\phantom{0}0.48$ &$\phantom{0}0.54$ &$\phantom{0}0.60$ &$\phantom{0}0.66$ &$\phantom{0}0.72$ & \textit{C1} & \textit{C2} \\
			\hline &&&&&&&&& \vspace{-0.3cm} \\
			computation time (s) & $\phantom{0}0.07$ & $\phantom{0}0.14$ & $\phantom{0}0.15$ & $\phantom{0}0.41$ & $\phantom{0}0.43$ & $\phantom{0}0.41$
			& $\phantom{0}0.42$ & $\phantom{0}0.43$ & $\phantom{0}0.43$ & 
			$\phantom{0}5.33$ & $\phantom{0}3.81$ \\
			\hline
			 &&&&&&&&& \vspace{-0.3cm} \\
			$R^{\delta}(\ds)$ & $\phantom{0}0.45$ & $\phantom{0}0.70$ & $\phantom{0}0.73$ & $\phantom{0}0.60$ & $\phantom{0}0.68$ & $\phantom{0}0.71$
			& $\phantom{0}0.75$ & $\phantom{0}0.75$ & $\phantom{0}0.78$ & $\phantom{0}0.61$ & $\phantom{0}0.71$\\
			balancing energy (kWh) & $\phantom{0}5.82$ & $\phantom{0}3.61$ & $\phantom{0}3.47$ & $\phantom{0}4.39$ & $\phantom{0}3.56$ & $\phantom{0}3.10$
			& $\phantom{0}2.86$ & $\phantom{0}2.79$ & $\phantom{0}2.66$ & $\phantom{0}4.81$ & $\phantom{0}3.56$\\
			\hline &&&&&&&&& \vspace{-0.3cm} \\
			cost $\ds$ (\euro) & $\bold{\phantom{0}4.86}$ & $\phantom{0}6.40$ & $\phantom{0}6.73$ & $\phantom{0}5.48$ & $\phantom{0}5.87$ & $\phantom{0}6.29$
			& $\phantom{0}6.68$ & $\phantom{0}6.84$ & $\phantom{0}6.96$ & $\phantom{0}5.40$ & $\phantom{0}6.24$\\
			cost $\{\Delta p_g(k)\}_{k \in \mathcal{K}}$ \textit{C1} (\euro) & $\phantom{0}4.06$ & $\phantom{0}2.57$ & $\phantom{0}2.49$ & $\phantom{0}3.07$ & $\phantom{0}2.51$ & $\phantom{0}2.19$
			& $\phantom{0}2.02$ & $\phantom{0}1.98$ & $\bold{\phantom{0}1.90}$ & $\phantom{0}3.41$ & - \\
			cost total \textit{C1} (\euro) & $\phantom{0}8.92$ & $\phantom{0}8.97$ & $\phantom{0}9.22$ & $\phantom{0}8.55$ & $\bold{\phantom{0}8.38}$ & $\phantom{0}8.48$
			& $\phantom{0}8.70$ & $\phantom{0}8.83$ & $\phantom{0}8.86$ & $\phantom{0}8.81$ & - \\
			cost $\{\Delta p_g(k)\}_{k \in \mathcal{K}}$ \textit{C2} (\euro) & $27.77$ & - & - & $19.86$ & $16.56$ & $14.46$
			& $13.53$ & $13.43$ & $\bold{13.10}$ & - & $17.51$ \\
			cost total \textit{C2} (\euro) & $30.36$ & - & - & $25.34$ & $22.42$ & $20.75$
			& $20.21$ & $20.27$ & $\bold{20.06}$ & - & $23.39$ \\
			\hline 
		\end{tabular}
	\end{center}
	\vspace{-0.6cm}
\end{table*}

A household provided with a PV generator and a domestic battery is selected as test case, similar to \cite{Appino18a}.
The data of PV production and load consumption is retrieved from the freely accessible dataset provided by Ausgrid \cite{Ratnam15a}.\footnote{The dataset offers the time series of the load and PV generation profile of 300 Australian households with installed rooftop PV systems for the time frame of 01/07/10 to 30/06/13. 
The utilized data refers to household $109$. 
Notice that we employ for $\{p_l(k)\}_{k\in \mathcal{K}}$ the hourly-averaged profile of the real production/consumption, considering that the storage compensates for the zero-mean intra-hour variability.} 
The technical specifications of the battery comes from the catalog of a commercial producer.\footnote{www.tesla.com/powerwall}
Accounting only for the usable capacity, these are: $\Esmin = 0$ kWh, $\Esmax = 13.5$ kWh, $\Psmin = -5$ kW, $\Psmax = 5$ kW, $\mun = 5\%$.    
In the present paper, we consider day-ahead scheduling with scheduling horizon spanning midnight to midnight. 
The scheduling horizon is divided into $N_d = 24$ dispatch intervals and the dispatch schedule has to be computed before midday, i.e. $\Dtk = 1\text{h}$, $\kstart = 12 + k_0$.     
For sake of simplicity, we use time invariant coefficients in cost function \eqref{eq:cost_f}: $c_{2}^+ = 0.05 \frac{\text{\euro} \cdot \text{h}}{\text{kW}}$, $c_{2}^-= 0.05 \frac{\text{\euro} \cdot \text{h}}{\text{kW}}$, $c_{1}^+= 0.3 \frac{\text{\euro} \cdot \text{h}}{\text{kW}}$, $c_{1}^-=0.15 \frac{\text{\euro} \cdot \text{h}}{\text{kW}}$. 
These values represent a pricing policy incentivizing self-consumption and load leveling.

The simulations are carried out in MATLAB, employing standard open-source optimization tools developed in the systems and control community to solve the scheduling problems.
Specifically, we use CasaDi \cite{Andersson13b} with the IPOPT \cite{Wachter06}.  
All the computations have been performed using a PC with an Intel\textsuperscript{\textregistered} Core\textsuperscript{TM} i5-6400 CPU at 2.70 GHz and 8.00 GB RAM.

We simulate and compare the effect of three different scheduling techniques: (i) the Deterministic Forecast Scheduling (DFS), where the schedule is computed applying deterministic forecasts, i.e. $\rv{\Delta P}_l \equiv 0$; (ii) the Probabilistic Forecast Scheduling (PFS), cf. Problem \eqref{eq:pfs_opt_problem}; (iii) the Scenario Forecast Scheduling (SFS), cf. Problem \eqref{eq:opt_problem}. 
The performance of the PFS is assessed using different values for the security level ranging from 0.42, to 0.72.
The simulations cover five different weeks in the time frame going between 01/02/13 and 30/06/13. 
To the end of covering the effects of seasonal changes, these weeks are selected in different months. 
We consider two different pricing policies of the imbalances, $\textit{C1}$ and $\textit{C2}$, to examine the effects of the various scheduling techniques on the final operating cost. 
Specifically, in $\textit{C1}$ and $\textit{C2}$ the tariff of imbalances is twice as high and, respectively, ten times as high as the one of the DiS. 
Thereby, power excess and shortage count as purchased power. 
Table \ref{tab:simulation_results} provides an overview of the results obtained with the different schemes. 
To foster comparison we restate also the results reported in \cite{Appino18a}.

The required probabilistic forecasts for both power and energy are created using quantile regressions \cite{Fahrmeir13,Koenker05} based on a k-nearest-neighbor data-driven approach \cite{GonzalezOrdiano16}.\footnote{The  forecasting models are generated with the open-source MATLAB toolbox SciXMiner \cite{Mikut17}. The data from 01/07/10 to 01/12/12 is utilized for training the model. Please notice that all forecasts are only based on historical power time series, since the Ausgrid dataset does not contain weather forecast.} 
These quantile regressions predict the quantiles of $\rv{P}_l(k)$ and $\rv{\Delta E}_l(k)$.
In the PFS, the quantiles of $\rv{P}_l(k)$ are used to determine the interval $[\underline{p}_{l,\probP}(k),\overline{p}_{l,\probP}(k)]$ according to \eqref{eq:interval_Pl}, whereas the quantiles of $\rv{\Delta E}_l(k)$ are used to fit the parameters of two logistic functions whose sum is utilized as a description of $F_{\rv{\Delta E}_l(k)}(\Delta e_l)$.\footnote{The fitting is done using least-squares. 
Extensive numerical studies have shown that the choice of a logistic function with six parameters $[a_1 ... a_6]$, i.e. $F_{\rv{\Delta E}_l(k)}(\Delta e_l)=\frac{a_1}{1+e^{-a_2(\Delta e_l-a_3)}}+\frac{a_4}{1+e^{-a_5(\Delta e_l-a_6)}}$, is able to reproduce the skewness of the quantiles. 
Other choices, ex. hyperbolic tangent, arctangent or specific algebraic functions, have shown poor results.} 
While the power quantile regressions take only past generated power data as input, the energy forecasting models receive as input the integrated values of the powers' median regression \cite{Appino18a}. 
In addition, quantile regressions are applied also in the SFS, to generate the various power forecast scenarios. In this case the regressions predict the quantiles of the power value an hour into the future. 
Each scenario is created (i) by randomly selecting one of the predicted quantiles, (ii) by using it as input of the one-hour-ahead quantile regressions, and (iii) by repeating the first two steps for the length of the extended scheduling horizon. 
Furthermore, we apply the algorithm presented in \cite{Conejo10} to reduce the number of scenarios to $N_s = 30$ and assign a weight $\omega_i$ to each of them.

The average computation times required to solve the considered scheduling problems are reported in Table \ref{tab:simulation_results}. 
One can see that all three variants are solved within fractions of a second (DFS and PFS) or within a few seconds (SFS). 
Thus, the computational load does not appear to be an implementation barrier for any of the scheduling formulations. 

\begin{figure}[t]
	\centering
        \includegraphics[width=0.495\textwidth]{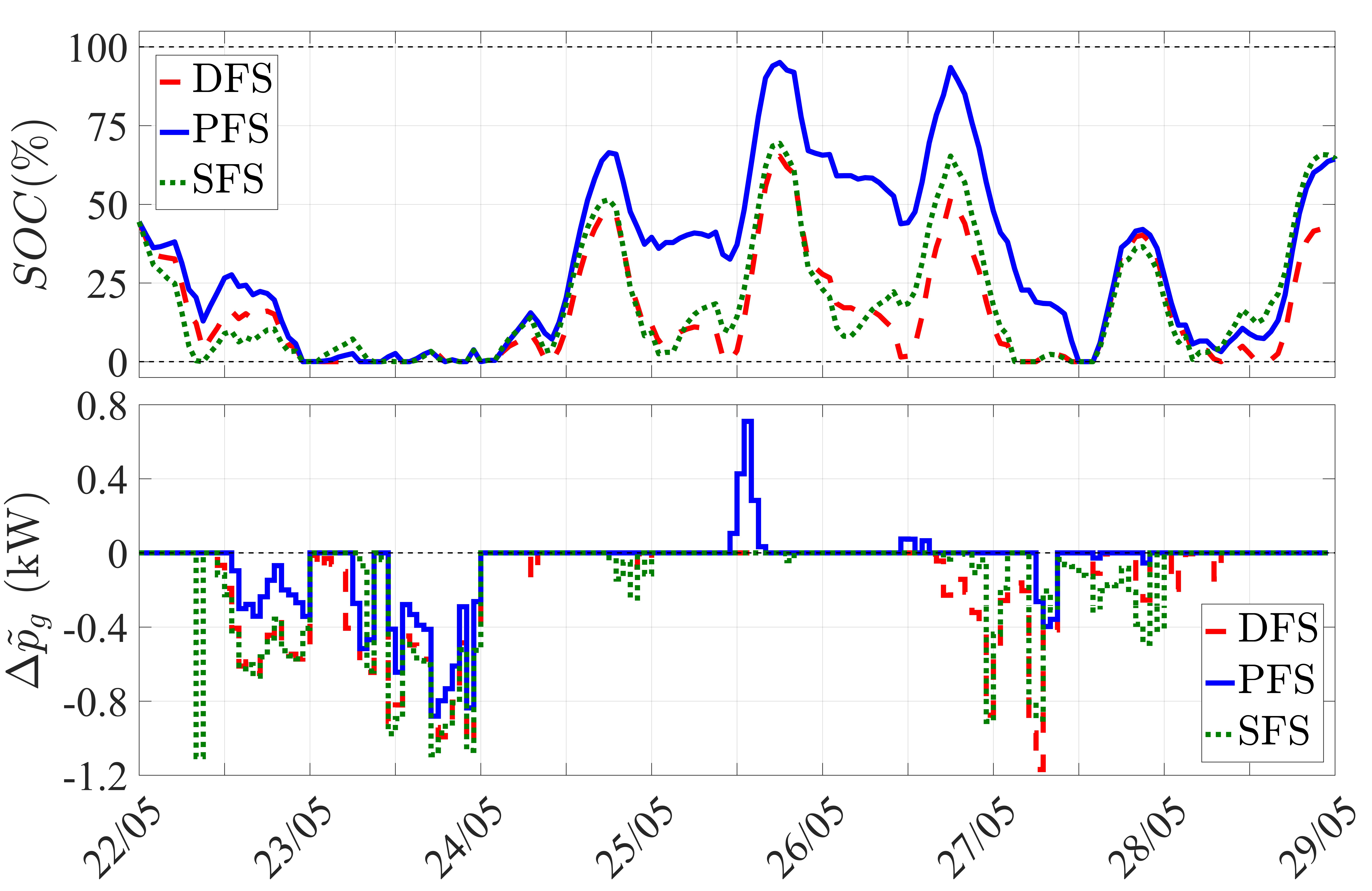}
    \caption{Comparison of results over a simulated week: state of charge and imbalances. Cost case \textit{C1}, PFS with $(1-\varepsilon) = 0.54$.}
    \label{fig:soc_imb_compare}
    \vspace{-0.5cm}
\end{figure}

To evaluate how well the DiS is met during the operation of the DF we define the scheduling tracking ratio 
\begin{equation*}
R^{\delta}(\ds)=\frac{\#\left\{ k \in \mathcal{K} \mid \left |  p_g(k) - \Pgsch{k} \right |\leq \delta \right\}}{\#\mathcal{K}},
\end{equation*}
where $\#$ denotes the cardinality of the set and $\delta = 10^{-4}$. 
The average values of $R^{\delta}(\ds)$ resulting from the different scheduling schemes are listed in Table \ref{tab:simulation_results}, as well as the average amount of energy required daily from the grid to compensate for the imbalances. 
This energy has to be considered as the total daily energy request, regardless of whether it was absorbed or injected into the grid.
The detailed imbalances profile for three different cases over a week is depicted in Figure \ref{fig:soc_imb_compare}.
Both from Table \ref{tab:simulation_results} and Figure \ref{fig:soc_imb_compare}, it can be noticed that the DFS has the worst tracking performance.
The PFS, instead, always achieves the desired outcome of meeting the security level, i.e. $R(\ds) \geq (1-\varepsilon_E)$.
In the SFS the tracking ratio depends on the pricing policy of the imbalances. 
This is aligned with the motivations behind the SFS, aiming at the best trade off between the DiS cost and the expected cost of imbalances.
In fact, as shown in Table \ref{tab:simulation_results}, better tracking performances are associated with the increment of the cost of the DiS. 

\begin{figure}[t]
	\centering
        \includegraphics[width=0.495\textwidth]{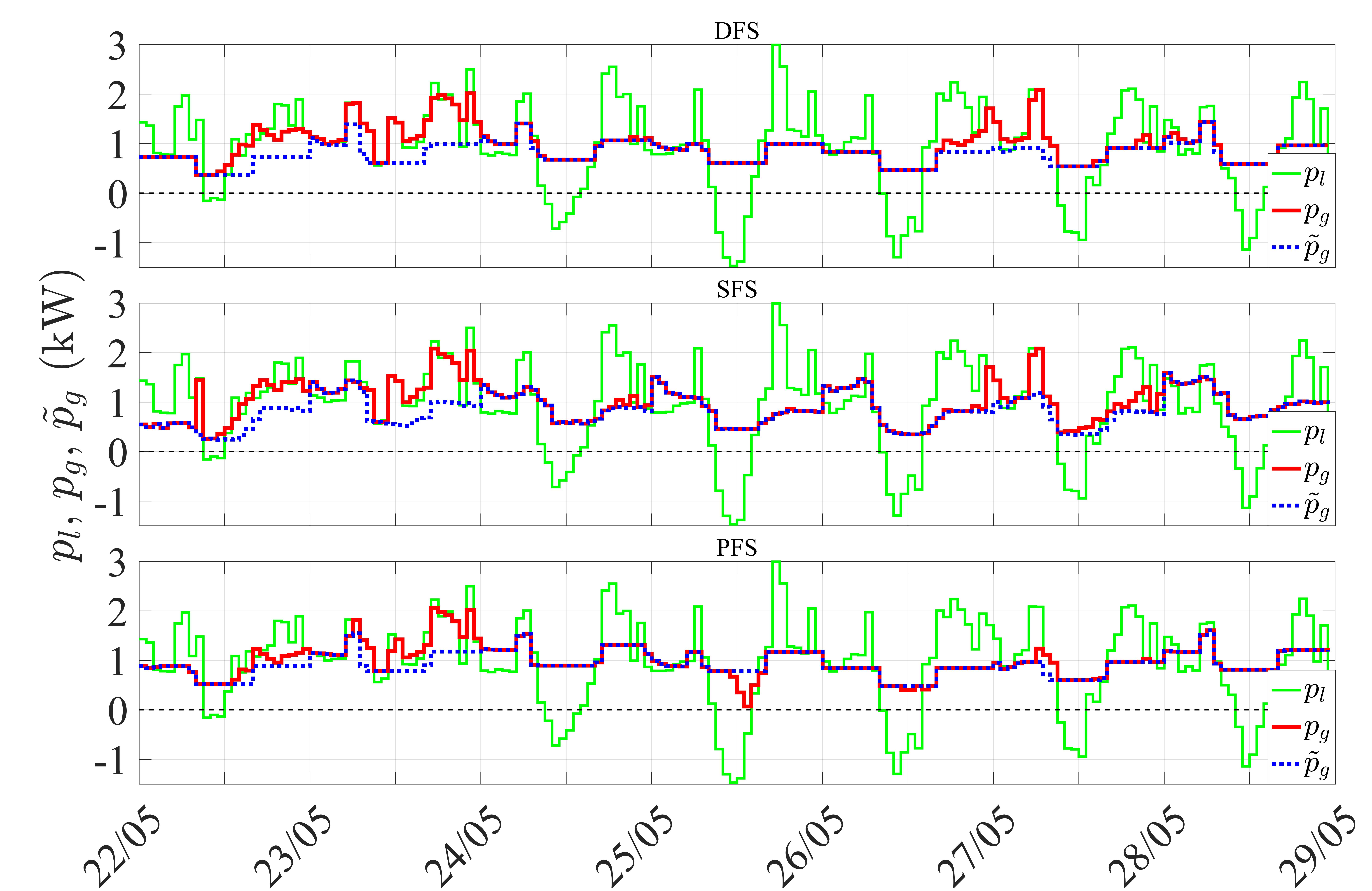}
        \label{fig:profile_dfs}
        \vspace{-0.3cm}
    \caption{Comparison of results over a simulated week: power profiles. Cost case \textit{C1}, PFS with $(1-\varepsilon) = 0.54$.}
    \label{fig:profile_all}
    \vspace{-0.55cm}
\end{figure}

These opposite tendencies are also visible between the different scheduling techniques. 
Figure \ref{fig:profile_all} reports the power output profiles over the same week applying different scheduling procedures. 
The green plot represents the baseline profile $\left\lbrace p_l(k)\right\rbrace_{k\in\mathcal{K}}$, the blue one represents the DiS $\ds$, and the red one the profile $\left\lbrace p_g(k)\right\rbrace_{k\in\mathcal{K}}$ resulting from the actual $\left\lbrace p_l(k)\right\rbrace_{k\in\mathcal{K}}$. 
While the DFS leads to the DiS with minimum cost, this DiS cannot be tracked as efficiently as the one computed using the PFS. 
The SFS tries to balance these two aspects.

However, while the SFS achieves a lower total cost compared to the DFS, the minimum actual total cost follows surprisingly from the application of the PFS with an appropriate security level in both pricing policies \textit{C1} and \textit{C2}. 
This phenomenon can be explained considering that the SFS has two limitations \cite{Fabietti16}: it optimizes the problem for a finite set of scenarios and it considers an unrealistic optimization of the imbalances.
These limitations do not affect the PFS, where the reserves to compensate for eventual imbalances are allocated on the basis of probabilistic forecasts for the energy profile. 
This way infinite possible realizations of $\{p_l(k)\}_{k\in \mathcal{K}}$ are considered without any assumption on an optimized redistribution of imbalances.
The improved allocation of reserves 
is also visible in Figure \ref{fig:soc_imb_compare}, which describes the State Of Charge (SOC) profiles for the different cases. 
It can be seen that the PFS leads to a complete exploitation of the storage capacity.
However, notice that the PFS does not outperform the SFS in minimizing the total operating cost for all the values of $(1 - \varepsilon_E)$.
Furthermore, the optimal value of $(1 - \varepsilon_E)$ to minimize the total cost differs with the pricing policy of the imbalances.
Still, a good value for $(1 - \varepsilon_E)$ can be computed by means of simulations. Therefore, the PFS can be efficiently applied to satisfy both the tracking requirements: limiting the number of imbalances and minimizing the total operating cost. 

Finally we remark that the PFS described in the present paper leads to better performance than the one presented in \cite{Appino18a}, because of the reduced conservativeness (cf. Section \ref{sec:security_level}). In particular: (i) higher security levels can be considered, (ii) the actual tracking is more aligned with the desired security level, and (iii) the total operation cost is reduced. 


\section{Conclusions}
\label{sec:conclusion}

The present paper investigated different techniques to compute an efficient schedule of the power exchange between a \textit{dispatchable feeder} and the utility grid. 
In particular, we propose a formulation of the scheduling problem that exploits probabilistic forecasts in terms of cumulative density function.
The result is a \textit{dispatch schedule} that can be tracked in operation with at least a given probability, called \textit{security level}.
We compare the proposed method to scheduling based on deterministic forecast and scenario-based optimization.
The simulation results show that the proposed method achieves an efficient computation of a dispatch schedule that not only ensures the desired security level, but that also leads to lower total operational cost than a scenario-based scheduling designed for total cost minimization.
Future work will consider scheduling of populations of storages and their coupling through distribution networks, extending the proposed approach to deferrable loads, and a proof-of-concept implementation.




\bibliographystyle{IEEEtran}
\bibliography{pscc18_bib}{}

\end{document}